# Vibe to Code: Elucidating Strategic Oscillation of Tacit Knowledge in Generative AI Design Workflows — An Exploratory Qualitative Study

Daisaku Sato

Graduate School of Knowledge Science, Japan Advanced Institute of Science and Technology (JAIST) / Sumitomo Mitsui Finance and Leasing Co., Ltd.

Corresponding author: dice310@gmail.com

## Abstract

The rapid adoption of generative AI tools has created new literacy demands for designers who must verbalize tacit knowledge through natural language prompts. Yet the micro-level cognitive processes by which designers externalize implicit intentions during iterative AI dialogue remain underexplored. This exploratory qualitative study examined five expert designers (11–20 years UI/UX experience, $M$ = 15.4 years) using think-aloud protocols. We identified “Strategic Oscillation”—experts’ intentional return to vague language (Vibe) after progressing toward operational specifications (Code), leveraging AI’s probabilistic nature for creative exploration. We observed shifts from “instruction” to “consultation” mode, deep reflection triggered by semantic collisions, and one participant’s disengagement revealing conceptual alignment as a tentative boundary condition. We propose the ECRT cycle (Expectation-Collision-Reflection-Transformation) operationalizing the SECI model’s “externalization” phase for human-AI collaboration. These findings yield three design patterns for metacognitive support and position AI as a “probabilistic mirror” catalyzing tacit knowledge externalization.

## 1. Introduction

### *1.1 Background: Prompt-Driven Design and Emerging Literacy Demands*

Generative AI tools have rapidly permeated design practice. Tools such as Figma AI, Replit AI, and Cursor now generate UI components directly from natural language text, fundamentally transforming how designers work. Whereas designers previously externalized mental images through sketches and wireframes, these tools position natural language “prompts” as the starting point for design.

This approach gained wide recognition when Karpathy (2025) coined the term “vibe coding” on social media. This approach—in which programmers and designers convey vague impressions (“vibes”) instead of specifying code or design details, delegating concretization to AI—can be situated within the scholarly discourse on end-user programming and natural language programming (Nardi, 1993; Lieberman et al., 2006). The implications are far-reaching: it represents a fundamental paradigm shift in the human–machine relationship.

*Scope of terminology:* In this study, “vibe coding (vibe to code)” is employed not as a transplantation of a programming technique from the software development context, but as an analytical concept describing the verbalization process by which designers move from vague impressions (vibe) to operational specifications (code). Accordingly, the findings of this study are not directly applicable to programming practice.

Under the previous paradigm (deterministic intent negotiation), humans fully defined the “correct answer” and machines faithfully executed instructions. Deviations from instructions constituted “errors”—noise to be eliminated. By contrast, collaboration with generative AI (probabilistic intent negotiation) involves AI probabilistically interpreting ambiguous human instructions and sometimes producing outputs that deviate substantially from expectations. Crucially, such deviations are not necessarily “failures.” Rather, they can function as “cues” for recognizing the ambiguity of one’s own instructions and refining verbalization.

### *1.2 Problem Statement: The Black Box of Verbalization Processes*

Polanyi’s (1966) famous proposition—“we can know more than we can tell”—captures a central tension in design expertise: much of what designers know resists verbalization. Schön (1983) conceptualized “reflection-in-action” and Cross (2004) identified “designerly ways of knowing,” both acknowledging that design expertise resists explicit verbalization.

Nonaka and Takeuchi’s (1995) SECI model theorized that dialogue and metaphor facilitate “externalization”—the conversion of tacit knowledge to explicit knowledge. However, their framework presupposed human-to-human dialogue. What happens when the “other” is an AI? As Collins (2010) noted, AI does not share tacit knowledge the way human interlocutors do. It lacks embodied experience and social context. Yet precisely for this reason, AI can function as a “probabilistic mirror” that objectively reflects one’s intentions. Existing research has primarily focused on prompt skill acquisition outcomes and AI collaboration efficiency. Studies that systematically trace the micro-level process changes—hesitations, rewrites, and realizations—occurring during dialogue remain limited.

### *1.3 Research Objectives and Questions*

This study aims to provide a microscopic description of how expert designers verbalize tacit knowledge and transform their strategies through iterative dialogue with generative AI. As an exploratory study oriented toward theory building rather than generalization, this study's contributions are threefold:

**Observational contribution:** Description of the "instruction-to-consultation" shift and lexical granularity oscillation (Strategic Oscillation) exhibited by expert designers.

**Methodological contribution:** Proposal of a protocol analysis framework combining ECRT with lexical granularity (L1–L3).

**Design implications:** Derivation of UI design requirements for metacognitive support in generative AI tools.

**Main Research Question (MRQ):** In the externalization phase of the SECI model, how does iterative dialogue with generative AI transform the process by which designers verbalize ambiguous images?

**SRQ1:** How do designers leverage gaps between expectations and outputs as cues for verbalizing latent intentions?

**SRQ2:** Through what lexical choices and rephrasings do designers linguistically reconstruct their instructions to AI?

**SRQ3:** How do the granularity of linguistic expression and aesthetic evaluation criteria structurally change through iterative dialogue?

This paper is organized as follows. Section 2 reviews related work and identifies research gaps. Section 3 presents the theoretical framework of the ECRT cycle and Strategic Oscillation. Section 4 describes the research design, participant selection, and data collection and analysis procedures. Section 5 reports three main findings, and Section 6 discusses theoretical and practical implications. Section 7 presents conclusions.

## 2. Related Work

### *2.1 Generative AI in Design and Prototyping*

The application of generative AI to the design domain is advancing rapidly. Koch et al. (2019) demonstrated the potential of positioning AI as a creative partner, while Yang et al. (2020) organized the challenges of AI utilization in UX design. A common characteristic of these tools is that users describe "what they want to create" in natural language, and AI interprets that intent to generate visual outputs. However, this "description of intent" itself poses a new cognitive challenge for designers.

**Limitations of existing research:** Most research on generative AI design tools has focused on output quality evaluation and usability assessment. Yet the micro-level cognitive processes of "how designers verbalize" and "what happens during the verbalization process" have not been sufficiently elucidated.

### *2.2 Human-AI Co-Creation and Prompting Strategies*

Knoth et al. (2024) demonstrated that prompt creation is a learnable skill dependent on AI literacy. Zamfirescu-Pereira et al. (2023) analyzed difficulties non-experts face in dialogue with LLMs and identified the fundamental challenge of "not knowing what to ask." Tankelevitch et al. (2024) systematically organized the metacognitive demands of generative AI use. Lee et al. (2025) showed, in a survey of 319 knowledge workers, that high trust in generative AI was associated with reduced critical thinking, while high task-specific self-efficacy was associated with enhanced critical thinking. Sidra and Mason (2025) developed and validated a "Collaborative AI Metacognition" scale, presenting important findings from the perspective of Kahneman's (2011) dual-process theory.

Simkute et al. (2024), writing in the *International Journal of Human–Computer Interaction*, identified the "ironies of generative AI." They analyzed the challenges of productivity decline due to user role shifts from "production" to "evaluation," inefficient workflow restructuring, and interruption, drawing on decades of human factors research. Their macro-level analysis provides a complementary perspective to the micro-level investigation in the present study.

**Limitations of existing research:** Most prompting research has focused on the "structure" and "techniques" of prompts. What remains underexplored is how expert designers verbalize tacit knowledge and what cognitive transformations they experience during the process remain insufficiently explored.

### *2.3 Design Expertise and Tacit Knowledge Externalization*

Polanyi's (1966) concept of tacit knowledge occupies a central position in design research. Nonaka and Takeuchi's (1995) SECI model theorized the conversion process from tacit to explicit knowledge (externalization). Externalization is facilitated through dialogue, metaphor, and analogy. And yet, the SECI model presupposed human-to-human dialogue, and the externalization process when the "other" is AI remains unexplored.

### *2.4 From Deterministic to Probabilistic End-User Programming*

The "vibe coding" phenomenon observed in this study can be theoretically positioned as a paradigm shift in end-user programming (EUP). Generative AI introduces a probabilistic EUP paradigm in which users function as "supervisors" or "curators," providing high-level intent (Vibe) while delegating logical implementation to the model's probabilistic nature. This shifts the core competency from "logic description" to "intent negotiation."

### *2.5 Research Gap and Positioning of This Study*

The above review of related work identifies the following gaps: (1) micro-level cognitive processes have not been systematically traced; (2) no study has operationalized the externalization phase of the SECI model in a human-AI collaboration context; and (3) the distinctive strategies exhibited by expert designers (Strategic Oscillation) have not been described.

*Literature collection method:* The literature for this review was collected using ACM Digital Library, Scopus, and Web of Science with the search string “generative AI” AND (“design” OR “tacit knowledge” OR “prompt*“) AND (”think-aloud” OR “externalization” OR “SECI”) for English-language publications from 2019–2025. Additionally, forward and backward citation tracking (snowballing) and foundational works in cognitive psychology and knowledge science (Polanyi, 1966; Nonaka & Takeuchi, 1995; Schön, 1983, etc.) were manually supplemented. This review is not intended as a systematic review but rather as a narrative review aimed at covering the conceptual scope necessary for constructing the theoretical framework.

## 3. Theoretical Framework

### *3.1 The ECRT Cycle Model: Theoretical Integration*

To operationalize the “externalization” phase of the SECI model for interactions with a probabilistic agent (generative AI), we propose the ECRT cycle. A central premise of this framework is that generative AI functions as a “probabilistic mirror”—a mechanism by which AI’s literal, non-empathic processing of language input returns probabilistically interpreted output that confronts users with the ambiguity of their own verbal expressions (this mechanism and its activation conditions are operationally defined and empirically examined in Section 6.1). This model integrates the following established constructs: Piaget’s accommodation (1952), Festinger’s cognitive dissonance (1957), Schön’s reflection-in-action (1983), and Kahneman’s dual-process theory (2011).

*Figure 1*. ECRT Cycle: Operationalizing SECI Externalization in Human-AI Collaboration.

Note: The Theoretical Roots panel also includes Kahneman's (2011) dual-process theory (see text).

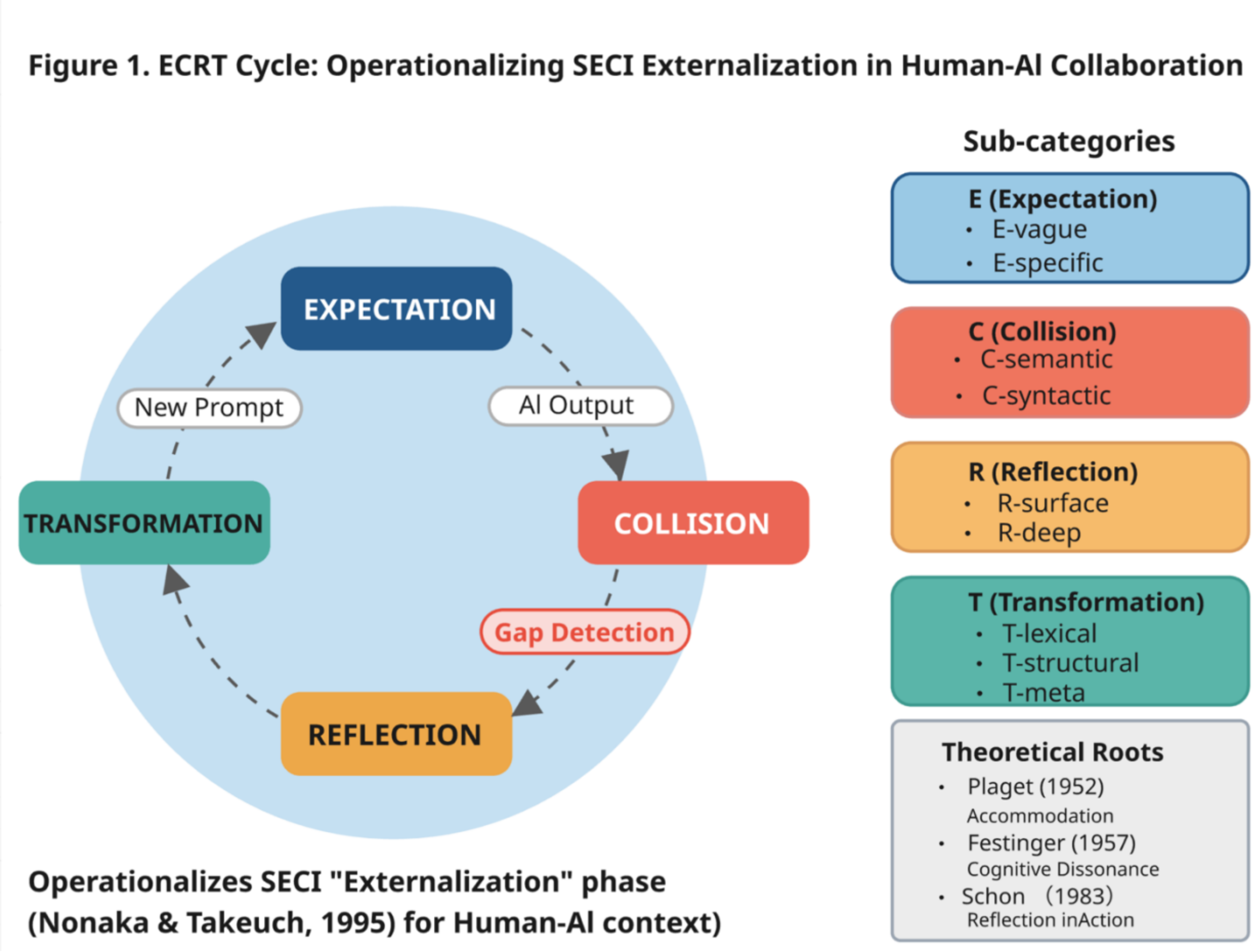

**Expectation (E):** The starting point of dialogue. Designers begin AI dialogue with ambiguous "images" or "feelings." These are further categorized into vague expectations (E-vague) and specific expectations (E-specific).

**Collision (C):** The perception of deviation from expectations when AI returns output. Semantic collision (C-semantic): the "meaning," "atmosphere," or "tone" of output does not match expectations. Syntactic collision (C-syntactic): the "structure," "layout," or "technical details" of output do not match expectations.

**Reflection (R):** Analysis of the cause of deviation. Surface reflection (R-surface): focus on superficial features of output such as "the color is different" or "the layout is different." Deep reflection (R-deep): focus on one's own cognitive processes, such as "my verbalization was ambiguous" or "I failed to share the concept."

**Transformation (T):** Change following reflection. Lexical transformation (T-lexical): changes in words or expressions. Structural transformation (T-structural): changes in prompt structure or sequence. Meta-transformation (T-meta): changes in strategy itself (e.g., shift from "instruction" to "consultation").

*3.1.1 Novelty of ECRT: Differences from Existing Reflective Models*

To clarify the theoretical positioning of the ECRT cycle, we conduct a systematic comparison with existing reflective learning models. Table 1 organizes differences across seven dimensions between Kolb's (1984) experiential learning cycle, Boud et al.'s (1985) reflective learning model, and Schön's (1983) reflection-in-action.

**Table 1. Comparison of the ECRT Cycle with Existing Reflective Models**

| Comparison Dimension | Kolb (1984) | Boud et al. (1985) | Schön (1983) | ECRT (This Study) |
|---|---|---|---|---|
| Reflection trigger | Concrete experience | Discomfort/perplexity | Unexpected results | **Collision with probabilistic output** |
| Dialogue partner | Self (introspection) | Self or others | Material/situation | **Probabilistic AI agent** |

| Comparison Dimension | Kolb (1984) | Boud et al. (1985) | Schön (1983) | ECRT (This Study) |
|---|---|---|---|---|
| Cycle directionality | Unidirectional/cyclical | Unidirectional/staged | Immediate/non-systematic | **Bidirectional oscillation** |
| Cognitive mechanism | Assimilation and accommodation | Affective reprocessing | Reflection-in-action | **Cognitive dissonance + dual-process switching** |
| Output predictability | Deterministic (teacher) | Deterministic (situation) | Partially predictable | **Inherently non-deterministic** |
| Scope of knowledge conversion | Individual learning | Individual learning | Professional practice | **Tacit → explicit knowledge (SECI externalization)** |
| Failure mechanism | Learning style mismatch | Affective barriers | Fixation on technical rationality | **Conceptual alignment failure → disengagement** |

*Note.* The distinctiveness of ECRT is summarized in three points: (1) collision with probabilistic AI as the reflection trigger, (2) description of bidirectional oscillation rather than unidirectional conversion, and (3) description of the co-occurrence of conceptual alignment failure and disengagement as a tentative boundary condition. These are not captured by existing models that presuppose human-to-human dialogue.

The core differences between the ECRT cycle and prior models are as follows. First, the non-deterministic nature of the reflection trigger. Kolb's "concrete experience" and Boud et al.'s "discomfort" arise from relatively predictable situations. In contrast, ECRT's "collision" is triggered by inherently non-deterministic AI output, which entails the possibility that identical prompts may generate different outputs. Second, the bidirectionality of the cycle. Existing models assume a unidirectional circulation of "experience → reflection → conceptualization → experimentation." ECRT describes Strategic Oscillation—experts' intentional return to ambiguity (L1)—which functions not as "regression" in existing models but as "strategic exploration." Third, the explicit specification of conceptual alignment as a boundary condition. P005's disengagement case suggested that conceptual alignment may be involved in the productive function of the ECRT cycle (though this is a tentative proposition based on a single case). This explicit specification of boundary conditions has not been attempted in existing models.

**Falsifiability of ECRT:** To ensure the falsifiability of the ECRT cycle, we specify the conditions under which this framework would not hold (falsification conditions). (1) Absence of reflection trigger: If AI output perfectly matches expectations (no collision occurs), stage C of ECRT is absent and the cycle does not initiate. This condition applies to work with deterministic tools (e.g., parametric CAD) or when AI output fully complies with user instructions. (2) Failure of the reflection capacity premise: For users without the habit or capacity for metacognitive reflection (e.g., novice designers), the progression from stage C to stage R may be impeded. (3) Failure of conceptual alignment: As demonstrated by P005's case, when users and AI cannot share basic task concepts, the ECRT cycle does not function productively and was observed to co-occur with disengagement. (4) Influence of language environment: The present data are from Japanese-language environments only, and the applicability of the L1–L3 lexical granularity classification itself has not been verified in environments with substantially different linguistic structures (e.g., isolating languages with fewer morphemes). These conditions make explicit that ECRT is not a universal cognitive model but a context-dependent framework that functions under specific preconditions.

### *3.2 Lexical Granularity Levels and Strategic Oscillation*

We classify designers' vocabulary directed at AI into three levels: L1 (Vibe)—abstract, impressionistic language such as "modern feel" or "sense of trust"; L2 (Semantic)—domain-specific vocabulary such as "hero section" or "minimalist layout"; L3 (Operational/Code)—concrete specifications such as "#000000" or "16px padding."

**Definition of Strategic Oscillation:** A dynamic oscillatory pattern in which expert designers do not progress unidirectionally from L1 (Vibe) to L3 (Operational) but intentionally return to ambiguous language levels (L1/L2) for creative exploration, then proceed again toward concretization (L3).

#### *3.2.1 Differentiation of Strategic Oscillation from Existing Concepts*

In proposing Strategic Oscillation (SO) as a novel concept, we conduct a systematic comparison with four established concepts in design research to clarify SO's discriminability.

**Table 2. Comparison of Strategic Oscillation with Existing Design Concepts**

| Concept | Commonality | Difference from SO | Observable Indicators | SO Identification Criteria |
|---|---|---|---|---|
| Divergent-convergent | Alternation between broad | SO involves movement between | Divergent-convergent: number of idea sketches | L1/L2 return after L3 attainment + |

| Concept | Commonality | Difference from SO | Observable Indicators | SO Identification Criteria |
| --- | --- | --- | --- | --- |
| thinking (Cross, 2004) | exploration and narrowing | lexical granularity levels (L1-L3), not divergence-convergence in the number of ideas. SO is triggered by collision with probabilistic AI output, not by the designer's intrinsic judgment | increases/decreases. SO: vocabulary in prompts returns from specific (L3) to vague (L1) | verbalization of intentional reason observed |
| Iteration (Schön, 1983) | Repetition of trial-feedback-modification | Iteration includes same-level modifications (e.g., L3→L3 fine-tuning). SO involves intentional cross-level movement (L3→L1), functioning as exploration restart | Iteration: same-granularity modification (changing color from #2563EB to #1D4ED8). SO: returning from L3 to L1 ("try a more trustworthy atmosphere") | Granularity level of modification target decreases (L3→L1/L2). Same-level modifications constitute iteration, not SO |

| Concept | Commonality | Difference from SO | Observable Indicators | SO Identification Criteria |
| --- | --- | --- | --- | --- |
| | | rather than error correction | | |
| Design co-evolution (Dorst & Cross, 2001) | Simultaneous development of problem and solution | Co-evolution involves parallel development of problem-solution spaces. SO involves oscillation between lexical granularity levels, which may not include changes to the problem definition itself | Co-evolution: redefinition of problem ("not LP but portal site"). SO: lexical level oscillation within the same problem | Lexical granularity changes while problem definition is maintained. Problem definition changes constitute co-evolution, which can coexist with but is distinguished from SO |
| Reflection-in-action (Schön, 1983) | Awareness and adjustment during action | Reflection is a cognitive process (internal). SO is an observable behavioral pattern (external). Reflection is part of the cognitive mechanism related to | Reflection: "Oh, my phrasing was ambiguous" (verbalization). SO: actual prompt granularity change from L3→L1 is executed (behavior) | In addition to cognitive awareness (reflection), actual prompt granularity change (behavior) is observed. Cases of reflection without SO (P005) exist |

| Concept | Commonality | Difference from SO | Observable Indicators | SO Identification Criteria |
|---|---|---|---|---|
| | | SO (stage R of ECRT) but is not SO itself | | |

*Note.* SO is not mutually exclusive with the above existing concepts and can coexist within the same session. SO's distinctiveness lies in the combination of three conditions: (1) movement between lexical granularity levels, (2) probabilistic AI output as the trigger for movement, and (3) level descent functioning as intentional exploration.

**Identification Rules for SO (Necessary and Sufficient Conditions):** SO is determined when all three of the following conditions are met.

**Condition 1 (L3 attainment premise):** At least one L3-level prompt must have been confirmed prior to the unit in question. Movement between L1/L2 without L3 attainment is distinguished as initial exploration, not SO.

**Condition 2 (Intentional return):** After L3 attainment, a prompt is issued that returns to L1 or L2 level vocabulary. Coders confirm that this return is not a "response to an AI error message" or "re-input due to system failure."

**Condition 3 (Verbalization of intentionality):** An intentional reason is expressed in the accompanying verbal data (e.g., "Let me reconsider from a broader perspective" or "Let me try approaching from the atmosphere"). If intentionality cannot be confirmed in the verbal data, SO is not determined.

**Supplementary behavioral indicators:** The following non-verbal behavioral patterns are referenced as supplementary indicators to support Condition 3 judgment (however, supplementary indicators alone do not satisfy Condition 3).

*(a) Consultation-mode shift:* Immediately following the L3→L1/L2 return, prompts requesting option generation appear (e.g., "Show me three variations" or "Suggest which is better"). Judgment criterion: structural change from directive (single output request) to interrogative or option-generation type.

*(b) Reframing:* After the return, re-declaration of purpose, tone, or concept is observed. Judgment criterion: the prompt contains expressions redefining design goals or constraints (e.g., "Since I was originally aiming for a ○○ atmosphere" or "Changing direction").

*(c) Prompt structural change:* The prompt structure qualitatively changes before and after the return. Judgment criterion: the pre-return prompt consists of operational instructions (specific numerical/parameter specifications), and the post-return prompt shifts to exploratory questions or abstract directional descriptions.

In the present data, the majority (7 of 9) of all 9 units determined as SO were accompanied by supplementary indicators (a) or (b).

*Measurement constraint:* Because the SO determination criteria include "verbalization of intentionality," SO may be underdetected when intention exists but is not verbalized. As think-aloud methods depend on participants' verbalization ability and metacognitive reporting habits, this point is acknowledged as a measurement bias in this study. Therefore, the SO frequency reported in this study potentially constitutes a conservative estimate (underestimate). However, the supplementary indicators (verbal content, score fluctuations) are themselves proxy measures, and the present design does not include means to directly observe the "true frequency" of SO.

*Handling of ambiguous cases:* When Conditions 1 and 2 are met but Condition 3 (verbalization of intentionality) is ambiguous, both coders independently judge and resolve disagreements through discussion. In the present dataset, this discussion occurred for 2 units, with 1 judged as SO and 1 as non-SO.

***Conditions under which SO does not hold (falsification and boundary conditions):*** To delimit the theoretical scope of SO, we specify conditions under which SO does not hold. (1) Failure of conceptual alignment: As demonstrated by P005's case, when users and AI cannot share basic task concepts, L1/L2 return after L3 attainment becomes "retreat due to confusion" rather than "strategic exploration restart," failing to meet SO Condition 3 (verbalization of intentionality). (2) Novices: Users who lack L3-level vocabulary structurally cannot meet Condition 1 (L3 attainment premise), making SO observation principally impossible. (3) Deterministic tool environments: When AI output is fully deterministic (identical input → identical output), "collision" with probabilistic output is unlikely, weakening the cognitive dissonance that triggers SO. (4) Language environment differences: This study targeted only Japanese speakers, and L1 (Vibe) level expression repertoires may depend on linguistic structure. The high-context nature of Japanese (see Section 6.4) may facilitate SO emergence, and emergence patterns in different language environments have not been verified. These boundary conditions make explicit that SO is not a phenomenon universally emerging across all users and environments but a context-dependent pattern observed when specific preconditions (expert-level vocabulary ability, probabilistic AI environment, minimum conceptual alignment) are met.

### *3.3 Methodological Positioning of Confidence-Satisfaction Scores: Structured Elicitation Device*

In this study, the 0–10 scores for confidence ("degree to which the prompt conveyed intent") and satisfaction ("satisfaction with output") are **not used as dependent variables for statistical analysis**. With N = 5 and 4–12 units per participant, the statistical pattern-detection power of the scores is insufficient, and such use would be inappropriate.

The function of the scores in this study is as a **structured elicitation device** complementing Ericsson and Simon's (1993) think-aloud method. The act of score assignment prompts participants to verbalize "why that number," thereby externalizing implicit evaluation criteria as verbal data. Accordingly:

**Primary data:** Verbal content during score assignment (e.g., "I gave it a 5 because the atmosphere is close but the layout is completely different")

**Contextual information:** The direction and magnitude of score changes are reported as descriptive indicators for contextualizing the intensity and content of verbalizations (e.g., "Accompanying a large negative gap from confidence 8 to satisfaction 1, P005 verbalized the following…")

**Not used:** Mean scores, variance, inter-participant statistical comparisons, correlation analyses

*All score reporting in this study is based on this positioning as a "structured elicitation device." Score changes reported in the results section (Section 5) serve to contextualize verbal data and do not rely on numerical patterns per se.*

### *3.4 Operational Definitions of Key Constructs*

Table 3 summarizes the operational definitions and coding judgment criteria for the key constructs used in this study. These definitions were refined through an abductive process, with initial theory-driven definitions undergoing data-driven modifications to reach their final form.

**Table 3. Operational Definitions and Coding Judgment Criteria for Key Constructs**

| Construct | Operational Definition | Inclusion Criteria (Code Assignment) | Exclusion Criteria (Code Non-assignment) |
|---|---|---|---|
| L1 (Vibe) | Sensory, impressionistic, non-domain-specific vocabulary. Does not contain design domain terminology | “Modern feel,” “sense of trust,” “cool,” “with a ○○ atmosphere” | Contains UI terminology (“minimal UI” = L2). Contains color codes or px values (L3) |
| L2 (Semantic) | Design-domain-specific professional vocabulary. Does not contain specific numerical values | “Hero section,” “CTA button,” “grid layout,” “utilizing white space” | Sensory expression only (L1). Contains specific CSS values or color codes (L3) |
| L3 (Operational) | Implementable concrete specifications. Contains numerical values, code, precise color specifications, etc. | “#2563EB,” “font-size: 16px,” “padding: 24px,” “border-radius: 8px” | Concept name only (“rounded button” = L2). No numerical values (L2 or below) |
| C-semantic | “Meaning/atmosphere/tone” of output does not match expectations. What-layer deviation | “Completely different atmosphere,” “This isn’t what I meant to convey,” “The tone is different” | Dissatisfaction with structure/layout only (C-syntactic). Reference to technical errors |
| C-syntactic | “Structure/layout/technical details” of output do not match expectations. How-layer deviation | “The layout is different,” “The button position is,” | Dissatisfaction with atmosphere/tone (C-semantic). Concept-level misalignment |

| Construct | Operational Definition | Inclusion Criteria (Code Assignment) | Exclusion Criteria (Code Non-assignment) |
|---|---|---|---|
| | | “It’s not two columns” | |
| R-deep | Reflection on one’s own cognitive processes/premises. Contains metacognitive references | “My phrasing was ambiguous,” “What was I actually trying to convey?,” “We weren’t sharing the premise” | Reference to output features only (R-surface). Modification instructions only (T phase) |
| Conceptual alignment | State in which human and AI share basic task concepts (e.g., “LP,” “hero section”) | Reference to the same concept using the same terminology, with output within the conceptual range. Operational modifications function | Different concepts referenced using the same terminology (e.g., P005’s “LP” recognition mismatch). Operational modifications consistently fail |
| Strategic Oscillation | Oscillatory pattern of intentional return from L3 to L1/L2, then proceeding again toward L3 | (1) After an L3-level prompt, (2) returns to L1/L2, and (3) verbalizes intentional reason | L1/L2 movement without L3 attainment. Re-input due to error. Unconscious expression changes |

*Note.* The distinction between C-semantic and C-syntactic was not theorized a priori but was inductively discovered from P001's data (Section 4.5). Conceptual alignment was added post hoc as a boundary condition from P005's disengagement analysis.

## 4. Methodology

### *4.1 Research Design*

*Epistemological stance:* This study draws on an interpretivist epistemology and adopts a constructivist ontology (Creswell & Poth, 2018). Designers' cognitive processes are understood as socially and contextually constructed; we do not assume a single objective reality measurable independently of context. Understanding participants' unique meaning-making and its generative processes thus aligns with the research objective. This epistemological stance motivates the selection of abductive analysis, first-person data collection through think-aloud methods, and interpretive coding.

This study was designed as an exploratory qualitative inquiry (Creswell & Poth, 2018), based on an abductive reasoning approach. The initial ECRT framework was deductively derived from existing theory and subsequently refined inductively through data analysis. This iterative oscillation between theory and data is characteristic of the abductive approach (Tavory & Timmermans, 2014) and is distinguished from purely deductive hypothesis testing and purely inductive grounded theory. Think-aloud protocols (Ericsson & Simon, 1993) were employed for data collection.

*Researcher positionality:* The first author is an insider researcher with 25 years of UI/UX design practice experience, belonging to the same professional domain as the participants. This position enhances sensitivity to designers' implicit verbalization patterns and subtle nuances of professional terminology, while entailing the risk of confirmation bias based on personal experience. Specifically, there is a danger that the first author may use personal verbalization habits as implicit interpretive standards. To mitigate this risk, the following procedures were implemented: (a) dual coding by an independent second coder from outside the HCI domain (Cohen's κ = 0.76–0.82), (b) a priori establishment of operational definitions—Table 3 definitions were finalized before coding commenced to minimize room for arbitrary judgment, and (c) proactive exploration of negative cases (P005's disengagement)—cases inconsistent with the theory were intentionally included as analysis targets to counteract confirmation bias.

### *4.2 Participants*

Selection criteria included: (1) 10+ years of UI/UX design practice experience, (2) lead designer experience on commercial products, and (3) 6+ months of professional use of generative AI tools.

**Table 4. Participant Demographics**

| ID | Years of Experience | Specialization | AI Use History | AI Use Frequency | Type | Outcome |
|---|---|---|---|---|---|---|
| P001 | 15 years | UI/UX, SaaS | 12 months | Daily | Leapfrog | Successful (satisfaction: 7) |
| P002 | 13 years | UI/UX, Healthcare | 8 months | 3–4 times/week | V-shaped recovery | V-shaped recovery (0→10) |
| P003 | 18 years | UI/UX, FinTech | 10 months | 3–4 times/week | Problem-solving | V-shaped recovery (2→9) |

| ID | Years of Experience | Specialization | AI Use History | AI Use Frequency | Type | Outcome |
|---|---|---|---|---|---|---|
| P004 | 20 years | UI/UX, Manufacturing SaaS | 12 months | Daily | Gradual improvement | Gradual improvement (3→6) |
| P005 | 11 years | UI/UX, Enterprise | 6 months | 2–3 times/month | Disengagement★ | Disengagement (satisfaction: 1×2) |

*Note.* ★P005 is the negative case (deviant case). ★ = confound candidate variable (P005 shows the lowest value and confounding with session behavior cannot be ruled out). AI use history = months of professional use of generative AI tools. AI use frequency = self-reported typical frequency confirmed in retrospective interviews. Satisfaction = final score (0–10, used as structured elicitation device, see Section 3.3). P005 showed the lowest values on three variables: years of experience (11 years), AI use history (6 months), and AI use frequency (2–3 times/month), and the possibility that these background differences influenced session behavior cannot be excluded (see Section 6.5.1).

**Table 5. Deviant Case Analysis: Contrast Between P005 and Successful Participants (P001–P004)**

| Comparison Dimension | P001–P004 (Successful) | P005 (Disengagement) |
|---|---|---|
| AI experience | 8–12 months ($M = 10.5$) | 6 months (minimum threshold) |
| AI use frequency | 3–4 times/week to daily | 2–3 times/month (lowest) |
| Design experience | 13–20 years ($M = 16.4$) | 11 years (lowest) |
| Strategy shift observation | Present (all 4 participants) | Absent (linear elaboration only) |
| Consultation-mode prompts | Spontaneously adopted | Not observed |

| Comparison Dimension | P001–P004 (Successful) | P005 (Disengagement) |
|---|---|---|
| Failure attribution style | Internal: “my verbalization” | External: “AI’s capability” |
| AI expectation setting | “Draft generator” / “starting point” | “It should understand me” |
| Conceptual alignment | Achieved in early session | Not achieved (“LP” mismatch) |
| Maximum negative gap | -5 (P002, recovered) | -7 (no recovery) |

*Note.* P005’s AI experience of 6 months (minimum selection criterion), AI use frequency (2–3 times/month, lowest among participants), and absence of strategy shift suggest the possibility that conceptual alignment requires a threshold of AI interaction experience (including not only use duration but also use density) (tentative proposition). Gap values descriptively indicate the direction of score changes as structured elicitation devices (see Section 3.3).

### *4.2.1 Sample Size Rationale*

The sample of N = 5 was strategically designed to maximize Information Power for theory building (Malterud et al., 2016). Following their five-dimension framework: (a) study aim—narrow and specific (micro-level cognitive processes during AI dialogue); (b) sample specificity—participants possessed an average of 15.4 years of professional design experience, constituting a “high-signal” group with dense, relevant knowledge; (c) established theory—the ECRT cycle and SECI model provide strong theoretical grounding that reduces the need for large samples to establish patterns; (d) quality of dialogue—the think-aloud protocol combined with retrospective interviews and structured elicitation yielded rich verbal data per participant; and (e) analysis strategy—abductive case-oriented analysis with explicit operational definitions. This combination yields high information power per participant. The sampling strategy also successfully captured a theoretically significant negative case (P005).

**Limitation:** The "disengagement" pattern is represented by P005 alone, and theoretical saturation has not been reached for this subcategory. Conclusions regarding the disengagement pattern are positioned as tentative propositions.

### *4.2.2 Trustworthiness*

Based on Lincoln and Guba's (1985) trustworthiness criteria, the following procedures were implemented: (a) Credibility: methodological triangulation—cross-referencing three data sources: think-aloud protocols, retrospective interviews, and confidence-satisfaction scores—and independent coding (Cohen's κ: ECRT subcodes = 0.76–0.82, lexical granularity = 0.89). (b) Transferability: detailed participant demographics (Tables 4 and 5). (c) Dependability: three-stage codebook development process (see Table 3 operational definitions). (d) Confirmability: detailed analysis of negative case (P005).

### *4.2.3 Ethical Considerations*

Prior to session commencement, all participants received comprehensive written and oral explanations of the following: (1) the objectives and procedures of the study, (2) data handling methods (including audio recording, screen recording, and anonymization procedures), (3) the voluntary nature of participation and the right to withdraw at any time without providing a reason, and (4) complete deletion of all data in the event of withdrawal. Written informed consent was obtained from each participant before the session began. The consent form explicitly specified that data use would be limited to academic research purposes, described data storage methods (encrypted storage), anonymization procedures (replacement of personally identifiable information with pseudonyms), and post-study data management policies. All participants signed the consent form after confirming their understanding of its contents.

The following data protection procedures were implemented. Session verbalizations were audio-recorded with participants' consent, and audio files were deleted upon completion of transcription. Screen-sharing recordings were used to document the chronological sequence of prompts and outputs, and anonymization processing was performed after analysis completion. Personally identifiable information in transcripts (names, company affiliations, project names, etc.) was replaced with pseudonyms. All data were stored in encrypted storage with access restricted to the research team. The consent form explicitly stated that participants could withdraw at any time without providing a reason, and that all data from withdrawing participants would be completely deleted. No compensation was provided for participation in this study.

Ethics approval statement: Formal ethics committee approval was not obtained for this study. The reason is as follows. At the time of data collection (January 2026), the author was affiliated with Sumitomo Mitsui Finance and Leasing Co., Ltd., a private-sector financial services organization that does not maintain an institutional review board (IRB) or ethics committee for non-medical behavioral research. Because no institutional ethics review mechanism was available, the author designed and conducted the study in strict accordance with the ethical principles of the Declaration of Helsinki (World Medical Association, 2013) and the Ethical Principles of Psychologists and Code of Conduct (American Psychological Association, 2017). The study posed minimal risk to participants: they voluntarily performed a routine professional design task (creating a landing page hero section) using commercially available tools in a familiar work setting, with no deception, no collection of sensitive personal data, and no physical or psychological intervention. Written informed consent was obtained from each participant prior to session commencement, documenting: (a) the objectives and procedures of the study, (b) data handling methods including audio and screen recording, (c) anonymization procedures, (d) the voluntary nature of participation, and (e) the right to withdraw at any time without consequence, with complete data deletion guaranteed upon withdrawal. Ethical oversight for subsequent phases of this research program will be provided by the institutional review board of the Japan Advanced Institute of Science and Technology (JAIST).

### *4.3 Task and Tools*

Participants designed the hero section of a landing page for a fictional B2B SaaS service. Tool: Figma Make (version 2025.11). The underlying generative model was not disclosed by the vendor. Data collection period: January 15–25, 2026.

The hero section task was selected because it requires designers to integrate multiple design dimensions—visual hierarchy, brand tone, typographic choices, and spatial composition—within a single, bounded artifact. This multidimensional nature naturally elicits vocabulary spanning all three granularity levels (L1 abstract/emotional, L2 structural, L3 technical), providing a suitable context for observing the verbalization of implicit design knowledge.

**Positioning of replicability regarding tool and model opacity:** This study used Figma Make (version 2025.11). The underlying generative model was not disclosed by the vendor and may change over time. Therefore, this study defines replicability not at the level of bitwise output replication but at the level of procedural and analytic replicability. Specifically, replicability is assessed by whether the proposed mechanisms (SO/ECRT cycle) can be re-observed in similar generative AI environments using the same protocol and coding scheme. To this end, we provide a complete codebook (Table 3), unit segmentation rules (Section 4.5), anonymized coding matrices (Tables 8–9), and reliability computation scripts (Section 6.6) to ensure auditability.

**Table 6. Fixed and Non-Fixed Elements of the Experimental Environment**

| Category | Element | Details/Notes |
| --- | --- | --- |
| **Fixed** | Tool | Figma Make version 2025.11 (same version across all sessions) |
| | Task | B2B SaaS LP hero section (same brief for all participants) |
| | Session duration | 30-minute AI dialogue (uniform across all participants) |
| | Prompt input method | Text input only (no voice or image input) |
| | UI settings | Default settings (no custom presets or templates) |
| | Data collection period | January 15–25, 2026 (11 days) |
| | Language | Japanese prompts only |
| **Non-fixed (uncontrollable)** | Underlying generative model | Not disclosed by vendor (Figma Inc.). Model type, version, and update timing not controllable by researchers |
| | Temperature parameters, etc. | Uncontrollable. Figma Make UI has no model parameter adjustment functionality |
| | Probabilistic output variation | Identical prompts may generate different outputs (see mini-verification below) |

*Note.* Non-fixed elements constitute the primary threat to this study's replicability. That said, since the object of analysis is not AI output quality but designers' cognitive responses to output

(ECRT cycle), output variation does not directly compromise analytical validity (elaborated in Section 6.5).

### *4.3.1 Output Variation Verification*

To evaluate the impact of underlying model non-fixedness on analytical conclusions, a mini-verification was conducted after data collection completion. Three representative prompts used during P001's session were re-executed three times each using the same tool and settings, and output variation was assessed.

**Verification results:** (1) Layout structure (number of columns, section arrangement) was generally consistent in 3 of 3 trials. (2) Specific color and typographic choices differed in 2–3 of 3 trials. (3) "Atmosphere" and "tone" similarity were evaluated based on the following rubric.

*Tone similarity rubric:* Three evaluation axes—(i) trustworthiness, (ii) approachability, (iii) modernity—were each rated on a three-level scale (low/medium/high), with "similar" defined as matching the original output on 2 or more of 3 axes. The first author and second coder independently evaluated the results, and in the majority (7 of 9) of 3 prompts × 3 trials = 9 comparisons, both evaluators judged "similar" (agreement: 7/9). This verification is exploratory and does not constitute proof of robustness. The 2 disagreements both involved the "approachability" axis.

**Implications for conclusions:** Probabilistic output variation primarily appeared at the L3 level (concrete specifications: color, font size, etc.), while variation at the L1 level (atmosphere/tone) was relatively small. The central findings of this study—Strategic Oscillation (movement patterns between lexical granularity levels) and the ECRT cycle (cognitive response patterns)—analyze cognitive processes in response to output rather than specific output content, and the possibility of L3-level variation overturning conclusions was not observed within this verification scope. However, this verification is limited and exploratory (n = 9; 3 prompts × 3 trials) and was not designed for systematic evaluation of output variation. We do not claim that output robustness has been demonstrated by this result. Systematic replicability verification remains a task for future research.

*4.3.2 Prompt-Output Log Examples (Anonymized)*

Below, we present the prompt and output summary for P001's Unit 3 (in which C-semantic collision → R-deep reflection was observed).

**Prompt (P001-U3):** *"Create a hero section for a modern corporate site that conveys trust but isn't too rigid. I want blue tones and a CTA button that stands out."*

**AI output summary:** A corporate-style layout with a dark navy background and white text. The CTA button was in fluorescent green. The overall impression was close to "an official financial institution website," diverging from P001's intended tone of "modern but approachable."

**P001's verbalization (think-aloud):** *"Hmm, this isn't right. Too rigid. There's trust, but it's not modern… Actually, what did I mean by 'modern'? Maybe I wasn't clear about that myself."*

**Coding:** C-semantic (tone mismatch) → R-deep (awareness of ambiguity in own verbalization) → T-lexical (changed to "SaaS startup style" in next prompt)

The above is a representative example of C-semantic collision → R-deep reflection. Below, we present an additional SO example and P005's consecutive disengagement unit sequence to ensure representativeness of examples.

**Example 2. P002-U8→U9 (SO example: large-amplitude oscillation)**

**Prompt (P002-U8):** *"Make it a corporate, trustworthy layout. Keep the navigation simple."*

**AI output summary:** A formal layout based on corporate navy and white. A polished impression, but the tone diverged from the "approachability" P002 had been exploring.

**P002's verbalization (think-aloud):** *"Yeah, this isn't bad per se… but what was the 'atmosphere' of this page supposed to be? I was aiming for something more friendly, but somehow I got pulled toward the rigid direction."*

**Coding:** C-semantic (overall tone divergence from intent) → R-deep (re-examination of the goal itself: "what was…supposed to be?") → T-meta (intentional return from L3→L1, strategy change). SO judgment: Condition 1 (L3→L1 return) ✓, Condition 2 (L1 return confirmed in verbalization) ✓, Condition 3 ("was aiming for something more friendly" = expression of intentional reason) ✓.

**Example 3. P005-U3→U4→U5 (Disengagement sequence)**

**Prompt (P005-U3):** *"Make the font a bit bigger, darken the header color a bit, and round the button corners."*

**AI output summary:** Slight font size increase, darker header color. But button corner rounding was not applied, and the overall layout balance differed from P005's expectations.

**P005's verbalization (think-aloud):** *"Even after explaining this much, it still doesn't work. Do I have to be even more specific?"*

**Coding:** C-syntactic (partial non-implementation) → R-surface (dissatisfaction with AI's comprehension = external attribution) → T-lexical (added further detailed instructions)

**Prompt (P005-U4):** *"Round the button corners with radius 8px. Make the header #2C3E50. Font size: body 16px, heading 24px."*

**AI output summary:** The specified values were largely reflected, but the overall "atmosphere" diverged further from P005's implicit image.

**P005's verbalization (think-aloud):** *"The numbers are right, but something's off. I don't know how to explain it anymore."*

**Coding:** C-semantic (overall impression mismatch despite numerical compliance) → R-surface ("don't know" = attribution fixation, no reflection on own verbalization) → Disengagement (assigned lowest score in next unit, expressed session termination intent)

P005's three-unit sequence demonstrates a pattern in which, despite continued operational detailing (L2→L3 direction), concept-level divergence remained unresolved, and disengagement was reached without consultation-mode shifts or R-deep occurring. The contrast with P001-U3 (Example 1) and P002-U8 (Example 2) illustrates how the success or failure of conceptual alignment co-occurs with subsequent cognitive processes.

### *4.4 Data Collection Procedure*

Each session lasted approximately 60 minutes per participant: introduction (5 min), practice time (10 min), image sharing (5 min), AI dialogue session (30 min), and retrospective interview (10 min).

*Note: No formal pilot study was conducted prior to data collection. The 10-minute practice phase within each session functioned as familiarization with the think-aloud protocol and AI tool, but did not constitute systematic pre-testing of the research protocol, codebook, or scoring procedures. Future replications should incorporate a dedicated pilot phase to identify potential ambiguities in the coding scheme.*

### *4.5 Analysis Method*

**Unit segmentation rules:** The unit of analysis was defined as one interaction unit consisting of one prompt submission → AI output generation → verbalization during output review.

**Segmentation criteria:** (1) A new prompt submission marks the unit boundary. (2) Multiple verbalizations regarding the same output are consolidated into one unit. (3) The point at which the participant assigns a score marks the unit endpoint.

**Consolidation criteria:** (1) Immediate additions to the preceding prompt (within 5 seconds), such as "Oh, actually, add this too," are consolidated into the same unit. (2) Re-submissions of the same prompt due to system errors are consolidated into the same unit.

**Exclusion criteria:** (1) Units during practice sessions (16 units). (2) Units in which output was not generated due to system errors.

**Example of a borderline case:** *In P002's Units 7-8, P002 muttered "a bit more…" while modifying and re-submitting the prompt. As the interval from modification to re-submission was 3 seconds, these were consolidated into the same unit based on the 5-second rule (Consolidation Criterion 1). Both coders agreed on this judgment.*

Of 50 total units, 34 were included in analysis (P001: 6, P002: 12, P003: 7, P004: 4, P005: 5). P002's higher unit count (12) compared to other participants (range: 4–7) reflects P002's adoption of a rapid-iteration strategy involving shorter and more frequent prompts. Session duration was uniform at 30 minutes for all participants. This uneven unit distribution means that P002 contributes disproportionately (35.3%) to aggregate patterns reported in Tables 8 and 9. Readers should note that patterns described as "cross-participant" may be weighted toward P002's interaction style. To mitigate this, we report participant-specific breakdowns throughout and present representative contrasting trajectories in Figure 2.

**Codebook development (abductive approach):** (1) Theory-driven codes: ECRT cycle (E, C, R, T) and lexical granularity (L1, L2, L3) were pre-established based on theoretical definitions (see Table 3). (2) Data-driven codes: Subcategories were inductively extracted from P001's and P002's data. Notably, the initial framework did not include the distinction between C-semantic and C-syntactic, but this distinction emerged from P001's contrasting responses to "atmosphere" and "layout" mismatches. (3) Codebook refinement: Verified and revised with data from the remaining three participants. P005's data prompted the addition of "conceptual alignment" as a boundary condition (see the final definitions in Table 3).

**Inter-coder reliability computation procedure:** The first author (Coder A) and a second coder (Coder B: a doctoral student in HCI research with no prior exposure to the ECRT framework) independently coded all 34 units.

**Computation unit:** Presence/absence for each unit × each code (binary judgment). Since multiple codes may be assigned to a single unit, computation was conducted code-by-code.

**ECRT subcodes (9 categories):** Code-specific Cohen's κ = 0.76–0.82 (median 0.76). Of 34 units × 9 subcodes = 306 binary judgments, 274 agreed and 32 disagreed. Code-specific κ values fall within the Substantial–Almost Perfect agreement range (Landis & Koch, 1977).

**Lexical granularity (L1/L2/L3, mutually exclusive 3 categories):** Cohen's κ = 0.89. Of 34 judgments, 32 agreed and 2 disagreed. Almost Perfect agreement (Landis & Koch, 1977).

*Inter-coder reliability was evaluated for each distinct labeling system. Aggregating different coding systems (non-exclusive binary judgments and exclusive categorical judgments) into a single index could introduce arbitrariness; therefore, this study reports system-specific κ as the primary metric. Note that while the Landis and Koch (1977) criteria are widely referenced, κ interpretation is context-dependent, and these are used as reference standards rather than normative thresholds.*

For ECRT subcodes, pooled PABAK = 0.79 is reported as a complementary measure (Byrt, Bishop, & Carlin, 1993). PABAK is not an alternative to κ but a complementary index for evaluating the impact of prevalence imbalance on κ. In the present data, certain subcodes (E-specific, C-semantic, T-lexical) had higher occurrence frequencies than others (T-meta, T-structural), resulting in varying prevalence indices across codes (PI = 0.00–0.50). For low-prevalence codes (T-meta: 8/34), the divergence between κ (0.76) and PABAK (0.82) was large, while for high-prevalence codes (C-semantic: 21/34), the divergence was small (κ = 0.81, PABAK = 0.82). This difference reflects the typical influence of prevalence imbalance, and co-reporting with PABAK ensures the robustness of reliability evaluation.

**Table 7. ECRT Subcode Confusion Matrix (Main 4 Codes Excerpted, N = 34 Units)**

| Coder A  B | C-sem present | C-sem absent | R-deep present | R-deep absent | κ (code-specific) |
|---|---|---|---|---|---|
| C-semantic present | 20 | 1 | — | — | |
| C-semantic absent | 2 | 11 | — | — | 0.81 |
| R-deep present | — | — | 12 | 1 | |
| R-deep absent | — | — | 2 | 19 | 0.82 |

*Note.* Table 7 excerpts confusion matrices for the main 2 codes due to space constraints. Complete confusion matrices for all 9 subcodes are presented in Appendix A. κ values were computed code-by-code: C-semantic: 0.81, C-syntactic: 0.76, R-deep: 0.82, R-surface: 0.76, T-lexical: 0.76, T-structural: 0.81, T-meta: 0.76, E-vague: 0.76, E-specific: 0.76. Each κ value is verifiable by recalculation from the confusion matrices in Appendix A.

## 5. Results

*Score values reported in this section constitute contextual information as structured elicitation devices (Section 3.3) and serve to support interpretation of verbal data.*

### *5.1 Participant Overview*

Four of five participants (P001–P004) expressed satisfaction during sessions. In contrast, P005 assigned the lowest score (1) twice consecutively and stated "I feel like giving up." The contrasting lexical granularity trajectories of representative participants (P001 vs. P005) are shown in Figure 2.

### *5.1.1 ECRT Subcode Distribution*

Table 8 presents the participant-specific occurrence frequencies of ECRT subcodes across 34 analysis units. Since multiple subcodes can occur within a single unit, column totals do not equal the number of units.

**Table 8. ECRT Subcode Distribution (34 Analysis Units)**

| Subcode | P001 (6) | P002 (12) | P003 (7) | P004 (4) | P005 (5) | Total (34) |
|---|---|---|---|---|---|---|
| E-vague | 3 | 5 | 2 | 2 | 3 | 15 |
| E-specific | 3 | 7 | 5 | 2 | 2 | 19 |
| **C-semantic** | 4 | 8 | 3 | 2 | 4 | **21** |
| C-syntactic | 2 | 6 | 5 | 3 | 1 | 17 |
| **R-deep** | 3 | 6 | 3 | 1 | 0 | **13** |
| R-surface | 2 | 5 | 3 | 2 | 4 | 16 |
| T-lexical | 3 | 7 | 4 | 2 | 3 | 19 |
| T-structural | 2 | 4 | 3 | 1 | 2 | 12 |
| **T-meta** | 2 | 3 | 2 | 1 | 0 | **8** |

*Note.* Numbers in parentheses indicate the participant's unit count. Bold indicates subcodes central to this study's claims. P005's highlighted cells (showing 0) = subcodes not observed for that participant. The denominator for Table 8 is 34 analysis units (Section 4.5). Multiple subcodes co-occurring within a single unit causes column totals to exceed the unit count.

*5.1.2 Association Between Collision Type and Reflection Depth: Exclusive Category Analysis*

Before examining the association between C-semantic and R-deep, we clarify the difference in coding methods. Non-exclusive coding (Table 8) allows simultaneous assignment of C-semantic and C-syntactic to a single unit, while exclusive coding (Table 9) re-codes each unit into exactly one of four categories: semantic-only, syntactic-only, both, or neither. Since the denominator is ambiguous under the non-exclusive coding of Table 8, we present the exclusive re-coding analysis as the primary result.

**Table 9. Exclusive Collision Category × Reflection Depth (N = 34 Units, Primary Result; Denominator = Units per Category)**

| Collision Category (Exclusive) | R-deep present | R-deep absent | Unit count | R-deep co-occurrence rate |
|---|---|---|---|---|
| Semantic-only (C-sem present, C-syn absent) | 7 | 3 | 10 | **70.0%** |
| Syntactic-only (C-syn present, C-sem absent) | 1 | 5 | 6 | 16.7% |
| Both (C-sem present ∩ C-syn present) | 4 | 7 | 11 | 36.4% |
| Neither (C-sem absent ∩ C-syn absent) | 1 | 6 | 7 | 14.3% |
| **Total** | 13 | 21 | 34 | 38.2% |

*Note.* Each unit classified into one of four exclusive categories. Denominator is the unit count per category (total N = 34). No statistical tests were applied given N = 34; reported as descriptive trends. Percentages are computed from small cell counts (e.g., 7/10, 1/6) and are statistically

unstable; a shift of one or two units would substantially alter the reported proportions. These figures serve as heuristic indicators for contextualizing verbal data patterns, not as evidence of stable population parameters.

**Interpretation of primary results:** Under exclusive categorization, semantic-only collisions co-occurred with R-deep at 70.0% (7/10), showing a notable difference from syntactic-only collisions (16.7%, 1/6). The Both category (36.4%) showed an intermediate value. This result suggests a tendency for semantic collisions (what-layer deviation) to be more strongly associated with deep reflection. However, causal relationships cannot be claimed from the present data, and this is interpreted as a descriptive pattern.

**Reference: Consistency check with non-exclusive analysis:** Under non-exclusive coding (Table 8), R-deep was simultaneously observed in 11 of 21 units (52.4%) where C-semantic collision was observed. For C-syntactic collision, R-deep was observed in 4 of 17 units (23.5%). The same trend was confirmed under both exclusive (Table 9) and non-exclusive (Table 8) analyses, indicating that this result is not an artifact of category overlap (both).

**Representative unit continuous context (P002-U4→U5→U6):**

**U4 (preceding):** P002 prompted at L2 level: "Place a large visual in the hero section." C-syntactic recorded for output (layout structure mismatch). R-surface verbalization: "The layout is wrong." → T-lexical: modified to "centered layout."

**U5 (focal unit: semantic-only + R-deep):** P002 returned from L2→L1: "Something warmer, more human." C-semantic recorded for output (tone mismatch). R-deep: "What does 'warmth' mean… I have a specific image in my head but can't put it into words." → T-meta: "Let me change my approach. Let me have AI suggest options" (strategy change).

**U6 (following):** P002 switched to consultation-mode prompt: "Show me three variations of warm corporate sites." Selected one from output: "This is close. I want to refine from here."

*Coding rationale:* U5's R-deep judgment was based on the verbalization "I have a specific image in my head but can't put it into words," which indicates reflection on one's own cognitive process (limits of verbalization) rather than output features. Both coders independently judged R-deep.

**Descriptive association between C-semantic and R-deep:** Under non-exclusive analysis (Table 8), R-deep was simultaneously observed in 11 of 21 units (52.4%) where C-semantic collision was observed. For C-syntactic collision, the rate was 4 of 17 units (23.5%). This trend was maintained under exclusive re-analysis (Table 9). This descriptive pattern suggests that semantic collision tends to be associated with deeper reflection, but causal relationships cannot be claimed from the present data.

**P005's distinctiveness:** P005 recorded C-semantic four times yet R-deep zero times (Table 8). R-deep was observed in 70.0% (7/10) of all semantic-only units (Table 9); notably, all seven R-deep instances belonged to P001–P004 (7/8 = 87.5%), whereas P005's two semantic-only units showed 0% R-deep co-occurrence (0/2). This contrast suggests that C-semantic collision does not automatically produce R-deep and that conceptual alignment may function as a mediating condition (though the alternative explanation of experience differences cannot be excluded; see Section 6.5).

### *5.2 Visualization of Strategic Oscillation: Cross-Participant Comparison*

*Figure 2.* Strategic Oscillation: Contrasting Trajectories of Expert Oscillators vs Linear Fixation. C1–C*n* denote successive prompt–response cycles (i.e., ECRT iterations) within each participant's session. The y-axis ("Level of Abstraction" in the figure) corresponds to the Lexical

Granularity construct (L1 = Exploration/Vibe, L2 = Semantic, L3 = Implementation/Code). Yellow ring = Strategic Retreat point; red × = disengagement/termination; green ✓ = successful convergence; “Conf” and “Sat” annotations indicate self-reported confidence and satisfaction scores.

**Figure 2. Strategic Oscillation: Contrasting Trajectories of Expert Oscillators vs Linear Fixation**

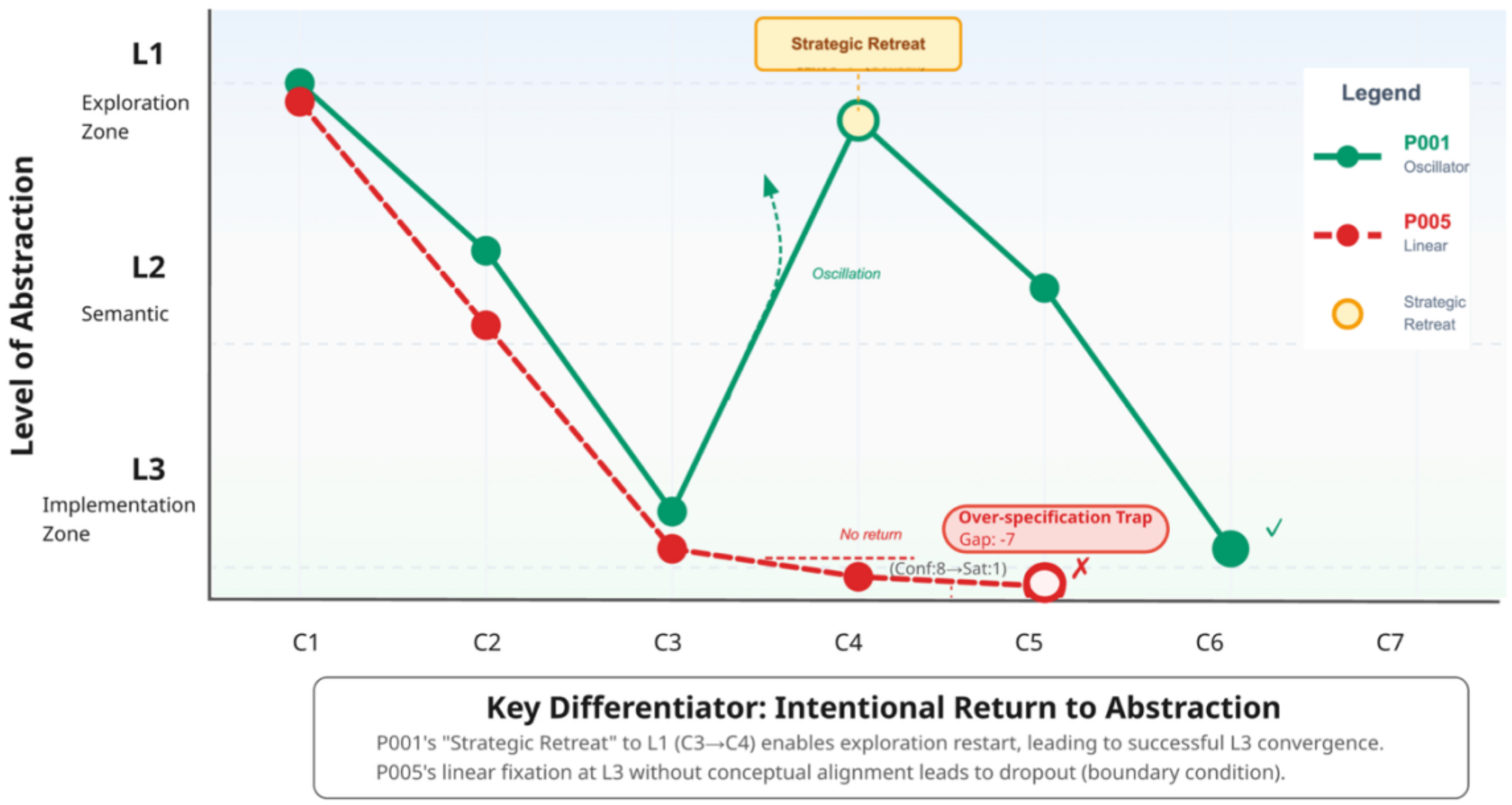

Figure 2 contrasts the lexical granularity trajectories of two representative participants, illustrating the core distinction between Strategic Oscillation and linear fixation. P001 (green line) exhibits the oscillatory pattern characteristic of expert designers: after progressing toward operational specificity (L3), P001 intentionally retreats to abstract, exploratory language (L1)—a “Strategic Retreat”—before re-converging toward L3. This intentional return to abstraction enables broader creative exploration and ultimately leads to successful convergence. In contrast, P005 (red line) follows a unidirectional descent from L1 toward L3 without any strategic return, representing “Linear Fixation.” P005’s trajectory shows no oscillation and terminates in disengagement.

### *5.3 Finding 1: Strategic Shift from “Instruction” to “Consultation” (Response to SRQ2)*

Expert designers not only refined instructions but also switched to consultation-mode prompting (optioning), retaining decision authority on the human side while broadening the exploration space.

**Dual-process theory interpretation:** From a dual-process theory perspective, the switch to consultation-mode prompting can be interpreted as a transition from System 1 automatic response (“just add more detail”) to System 2 strategic re-evaluation (“reconsider the strategy itself”). This type of strategy shift was not observed in P005, but this study did not directly measure System 1/System 2, so this explanation remains an interpretive framework.

**Boundary condition:** The switch to consultation mode was not observed in P005 (see Table 5). This suggests the possible involvement of conceptual alignment failure (see Table 3 operational definitions), but alternative explanations such as experience differences and AI use experience threshold effects cannot be excluded, making the boundary condition claim tentative.

Cross-participant comparison revealed diverse motivations for the consultation-mode shift. P001 achieved an early shift at Unit 2 of 6. After experiencing C-semantic collision in U2, P001 verbalized “It’s faster to have several options shown rather than deciding myself” and issued a clear consultation-mode prompt in U3: “Show me three patterns.” This verbalization indicates the moment P001 redefined AI’s role from “instruction executor” to “proposal generator.” P003 also showed a similar shift at U4, but through a different process. P003 verbalized “Hmm, I was trying to decide everything myself, but AI might have more options” and reached the shift through recognition of self-limitation. While P001 shifted from an efficiency perspective (“faster”), P003 adopted consultation mode as cognitive resource reallocation (“more options”). This contrast suggests that the motivation for consultation-mode shifts varies with designers’ cognitive framing.

P004 showed the most cautious approach. With 20 years of experience, P004 verbalized in early units “First, I want to accurately convey what’s in my head” and resisted transitioning to consultation mode. However, in U3, P004 realized “Fine-tuning each time doesn’t change things much,” and in U4, used a consultation-mode prompt for the first time: “Suggest two directions.” P004’s case shows that the “control orientation” based on long experience can serve as a barrier to the consultation-mode shift, while the accumulation of repeated expectation-output gaps ultimately functions as a shift trigger. This gradual transformation is consistent with the idea that transition from Kahneman’s System 1 to System 2 can occur gradually rather than instantly.

### *5.4 Finding 2: Qualitative Differences in Collision and Reflection Triggers (Response to SRQ1)*

Semantic collisions (C-semantic) tended to co-occur with deep reflection (R-deep), associated not with mere correction but with redefinition of “what I wanted to say.” Conversely, when conceptual alignment was lacking, collision tended to be associated with disengagement rather than reflection.

**Case 1—P002 (V-shaped recovery):** P002 assigned the lowest scores during the first three output reviews, verbalizing “It’s completely different from what I was thinking… I couldn’t verbalize it.” This is a clear case of simultaneous C-semantic collision and R-deep reflection. This progressed to “It’s better to have AI suggest options and I select rather than leaving it to AI” (T-meta: strategy change), and through the transition to consultation-mode prompting, P002 ultimately reached the highest score.

**Case 2—P005 (Disengagement):** P005 assigned a high confidence score and the lowest satisfaction score, and accompanying this gap, verbalized "Even after explaining this much, it still doesn't work." This is a case where C-semantic collision remained at R-surface without progressing to R-deep or T-meta. P005 then verbalized "I don't even feel like scoring anymore. I feel like giving up," showing a disengagement tendency.

*Methodological note:* References to scores above constitute descriptive information for contextualizing verbal content. The analytical focus is on the qualitative content of verbalizations accompanying score assignment, not on numerical patterns (see Section 3.3).

The chronological progression of collision types in P001 illustrates the staged nature of conceptual alignment and operational adjustment. Tracking P001's 6 units, C-semantic collisions were dominant in U1–U2 ("the atmosphere is completely different," "the tone is off"). From U3 onward, C-syntactic collisions increased ("column width," "button placement"). This progression pattern suggests a staged process in which designers first establish conceptual-level alignment, then transition to operational-level fine-tuning. Particularly noteworthy is P001's verbalization in U2: "What did I mean by 'modern'?" This verbalization was classified as R-deep, but it went beyond mere self-reflection to constitute a metacognitive act attempting to make explicit the aesthetic evaluation criteria held as tacit knowledge. Polanyi's (1966) "we can know more than we can tell" was experientially confirmed in this scene, triggered by AI dialogue.

P003 exhibited a distinctive "comparative reflection" pattern in 2 of 3 C-semantic collisions. Rather than simply contrasting AI output with personal expectations, P003 verbalized "This is interesting in its own way, but what I wanted to say was…" acknowledging the alternative value of AI output while re-confirming personal intent. This pattern was not observed in P001 or P002. P003's 18 years of experience may be associated with the flexibility to evaluate AI output as an "alternative" rather than an "error." However, this interpretation cannot rule out confounding between years of experience and design philosophy, and is reported as a descriptive observation. We examine in greater detail the mechanism by which P005 never exhibited R-deep. Tracking P005's verbal data chronologically reveals a consistent characteristic in responses to collision. U1: "It's different from what I expected" (R-surface), U2: "Maybe I should write in more detail" (T-lexical), U3: "Even after explaining this much, it still doesn't work" (external attribution), U4: "I don't know how to explain it anymore" (attribution fixation). This four-stage progression shows that P005 consistently maintained an external attribution frame of "the problem is with AI's comprehension ability." For R-deep to occur, a shift to internal attribution—"the problem is with my verbalization"—is necessary, and in P001–P004, C-semantic collision functioned as a trigger for this shift. In P005, conceptual alignment failure, external attribution frame fixation, absence of R-deep, and disengagement were simultaneously observed. These co-occurrence patterns suggest that conceptual alignment may have been mediating, but alternative explanations such as experience differences cannot be excluded, and interpretation remains tentative (see Section 6.5.1).

### *5.5 Finding 3: Strategic Oscillation of Vocabulary (Response to SRQ3)*

Lexical granularity did not increase monotonically; rather, an oscillatory pattern of context-dependent return to L1 was observed. This return functioned not as "regression" but as "exploration restart," and was associated with eventual L3 convergence. As shown in Figure 2, the oscillatory pattern (exemplified by P001) was associated with success or recovery, while P005's unidirectional pattern terminated in disengagement.

**Representative unit excerpts: Patterns determined as SO**

**Example 1 (P001-U4→U5, leapfrog SO):** In U4, P001 issued an L3-level prompt ("#2563EB, font-size: 48px, line-height: 1.2") and reviewed the output. In U5, P001 returned to L1 with "Something with more startup energy." Verbalization: "I specified too many numbers. Let me restart from the image." SO judgment rationale: Condition 1 (L3 attained) ✓, Condition 2 (L3→L1 return) ✓, Condition 3 (verbalization of intentionality: "restart") ✓.

**Example 2 (P002-U8→U9, V-shaped recovery SO):** In U8, P002 issued an L3-level prompt ("CTA button with border-radius: 24px, background #FF6B35"). In U9, P002 returned to L1: "Wait, what was the 'atmosphere' of this page again?" Verbalization: "I was too deep in the details and lost the overall feel." SO judgment rationale: All 3 conditions met.

**Example 3 (P003-U5→U6, problem-solving SO):** In U5, P003 reached L2→L3 ("12-column grid"). In U6, P003 returned to L1/L2 boundary: "A more calm, professional impression." Verbalization: "The grid is right, but I need to settle the overall air first." SO judgment rationale: All 3 conditions met.

**Patterns not determined as SO (counterexamples)**

**Counterexample 1 (P004-U2→U3, same-level modification):** In U2, P004 specified "padding: 32px" (L3). In U3, P004 modified to "change padding to 24px" (L3). This is an L3→L3 same-level modification that does not meet Condition 2 (level descent). Judged as iteration.

**Counterexample 2 (P005-U3→U4, movement without intentionality):** In U3, P005 used "make the font bigger" (L2/L3 boundary). In U4, P005 moved to "something just… better" (L1). Still, the verbalization was "I don't know what to do anymore"—an expression of confusion rather than intentional exploration. Condition 3 (verbalization of intentionality) not met; not judged as SO.

In a qualitative analysis of SO, we examined oscillation amplitude patterns. All 9 SO units were observed among P001–P004 (P001: 2, P002: 3, P003: 2, P004: 2; P005: 0), consistent with the absence of SO Condition 3 (verbalization of intentionality) in P005's verbal data. Classifying these 9 units by amplitude (level difference of return destination) revealed two patterns. "Large-amplitude SO" (L3→L1; 5 units) involved fundamental re-examination of direction, with design concept changes observed in subsequent prompts. In P002's U8→U9 (large-amplitude SO), following the L1 return ("What was the 'atmosphere' of this page again?"), the output direction qualitatively shifted from "corporate/solid" to "approachable/casual." "Small-amplitude SO" (L3→L2; 4 units) involved re-examination at the domain-specific vocabulary level for specific elements. In P003's U5→U6 (small-amplitude SO), following the L2 return ("a more calm impression"), the change was limited to the color palette while the overall structure was maintained. This contrast is a descriptive pattern suggesting that SO amplitude may be associated with the scope of subsequent transformation.

P004 exhibited a qualitatively distinct "gradual oscillation" pattern across four cycles. In each of P004's cycles, the amplitude gradually decreased (L3→L1→L3→L2→L3→L2→L3). P004 verbalized "Gradually, the image in my head became clearer," indicating that oscillation reflected gradual convergence of the exploration space. In contrast to P001's and P002's "large-amplitude oscillation → rapid convergence" patterns, P004 systematically and gradually reached final L3 convergence. Furthermore, P004 stated in the retrospective interview "Rather than deciding details from the start, going back and forth several times to gradually improve accuracy suits me better," suggesting that SO had been internalized as a conscious strategy. This "controlled exploration" underpinned by 20 years of experience is qualitatively different from the more intuitive and improvisational SO of P001 and P002, indicating that individual differences exist even at the expert level.

### *5.6 Integrated Response to MRQ*

Dialogue with generative AI reorganizes externalization not as a unidirectional "ambiguous → concrete" conversion but as an oscillatory movement (Strategic Oscillation) that maintains conceptual alignment. AI output functions as a "probabilistic mirror," triggering reflection through collision, and co-occurring with vocabulary and strategy transformation. Within the scope of the present data, this transformation was observed in situations where conceptual alignment was established to a certain degree. Meanwhile, alignment failure and disengagement were simultaneously observed in P005, but additional data are needed to generalize this as a boundary condition, and it is treated here as a tentative proposition.

## 6. Discussion

### *6.1 AI Functions as a Probabilistic Mirror*

Our findings point to a distinctive mechanism by which generative AI engages with the "externalization" phase of the SECI model. Because AI lacks empathy and contextual understanding, it takes designers' verbal expressions "at face value" and returns probabilistically interpreted results. This "probabilistic mirror" function forced designers to confront their own linguistic ambiguity—and in doing so, catalyzed reflection. This is consistent with Festinger's (1957) cognitive dissonance theory.

The operational definition and observational indicators of the "probabilistic mirror" are specified as follows. The probabilistic mirror refers to a mechanism by which generative AI (a) processes user language input literally, without empathic inference, (b) returns probabilistically interpreted output, and thereby (c) generates situations in which users confront the ambiguity or implicit assumptions in their own verbal expressions. As observational indicators, the co-occurrence pattern in which R-deep (reflection on one's own cognitive processes) follows C-semantic collision (mismatch of output tone/atmosphere) in the present data (Table 9: R-deep co-occurrence rate of 70.0% in the semantic-only category) indicates the activation of this mechanism. In contrast, when conceptual alignment was not established (P005), the same C-semantic collision remained at external attribution (R-surface), and the mirror function did not activate. Accordingly, the probabilistic mirror is presented not as a poetic metaphor but as a description of an operationally definable cognitive mechanism, constituting a testable proposition that includes its activation conditions (minimum establishment of conceptual alignment) and activation indicators (C-semantic → R-deep co-occurrence).

This finding complements the "ironies of generative AI" identified by Simkute et al. (2024) in this journal. While they documented macro-level productivity decline, the present study's micro-level analysis reveals the underlying cognitive mechanisms: the probabilistic mirror forces a transition from System 1 production to System 2 evaluation, which is simultaneously disruptive (consistent with their "irony" framing) and generative (enabling tacit knowledge externalization).

### *6.2 The Paradox of Specificity*

The received wisdom is straightforward: more specific prompts yield better results. Our findings directly challenge this notion. Expert designers strategically alternated between "specificity" and "ambiguity" (Figure 2). P005's case suggests the proposition that "conceptual alignment precedes operational alignment."

### *6.3 Design Patterns and UI Requirements*

**Design Pattern 1: Conceptual Alignment Verification UI.** Problem: Users who input operational details before establishing basic concepts reach P005-type disengagement. Lee et al. (2025) showed that overreliance on generative AI suppresses critical thinking, and conceptual alignment failure may amplify this risk. Solution: An interaction pattern that confirms basic concepts before operational details.

**Design Pattern 2: Consultation Shift Nudge.** Problem: Users facing consecutive failures tend to repeat the same approach. Solution: A nudge encouraging a switch to consultation-mode prompting upon detecting consecutive negative gaps.

**Design Pattern 3: Lexical Granularity Level Feedback.** Problem: Users are unaware of the granularity level of their own prompts. Solution: Real-time feedback of lexical granularity level (L1/L2/L3) during prompt input.

**Validity assessment of design patterns:** The three design patterns above are design hypotheses derived from exploratory data and have not undergone systematic validity assessment prior to implementation. However, we organize the derivation rationale, design requirements, and anticipated failure conditions for each pattern and attempt a tentative validity assessment. DP1 (Conceptual Alignment Verification UI) was derived from the co-occurrence pattern of P005's disengagement and conceptual misalignment (Section 5.4). An anticipated failure condition is that the verification step may interrupt users' natural workflow and paradoxically promote disengagement. DP2 (Consultation Shift Nudge) was derived from the successful "instruction → consultation" shift patterns observed in P001–P004 (Section 5.3). An anticipated failure condition is that inappropriately timed nudges may violate user autonomy. DP3 (Lexical Granularity Feedback) is based on the hypothesis that consciousness of SO supports metacognition. An anticipated failure condition is that feedback may increase cognitive load and particularly inhibit free exploration at the L1 level. Empirical evaluation of these design patterns (e.g., prototype implementation and expert evaluation, usability testing) is positioned as a future research task (see Section 7.3).

### *6.4 Linguistic and Cultural Considerations*

All participants in this study were Japanese designers who prompted in Japanese. Japanese permits greater ambiguity than English through subject omission, context-dependent interpretation, and numerous hedging expressions (e.g., "~na kanji de" [with a ~ish feel]). This rich repertoire of ambiguous expressions may have been associated with the emergence of L1 (Vibe) level prompting.

*Scope limitation:* The association between the above linguistic features and the Strategic Oscillation observed in this study is merely a tendency observed within the present data scope (N = 5, Japanese speakers only). The causal claim that the high-context nature of Japanese facilitates SO cannot be derived from this study design. Whether SO is a pattern specific to Japanese speakers or a language-independent expert strategy requires cross-linguistic comparative research (see Section 7.3).

Communication norms rooted in high-context culture (Hall, 1976) may be associated with a tendency for designers to expect "intent" to be inferred from minimal cues. P005's frustration may partially reflect culturally specific expectations regarding the inferential capacity of communication partners. That said, multiple alternative explanations exist for P005's disengagement beyond linguistic and cultural factors (see Section 6.5 below).

### *6.5 Limitations and Boundary Conditions*

#### *6.5.1 Alternative Explanations for P005's Disengagement*

Against this study's interpretation of P005's disengagement as "lack of conceptual alignment," the following alternative explanations exist. Since these cannot be fully discriminated from a single negative case, we organize the refutability of each alternative.

**Table 10. Analysis of Alternative Explanations for P005's Disengagement**

| Alternative Explanation | Evidence in Present Data | Refutability | Judgment |
|---|---|---|---|
| Difference in years of experience | P005 had 11 years (lowest). P001-P004 had 13-20 years (*M* = 16.4) | The gap between P003 (18 years) and P005 (11 years) is large, but between P002 (13 years) and P005 (11 years) is small. Years of experience alone do not sufficiently explain disengagement | Partially refutable. However, cannot be fully excluded as a confound |
| Difference in AI use experience | P005 had 6 months, 2-3 times/month (minimum criteria). P001-P004 had 8-12 months, 3+ times/week | Possible threshold effect of AI experience. Whether the difference in use frequency (2-3 times/month vs. 3-4+ times/week) is qualitatively different from use duration (6 vs. 8 months) cannot be determined from the present data | Not refutable. Threshold effect remains a tentative proposition. Additional use frequency data suggest a compound experience difference |

| Alternative Explanation | Evidence in Present Data | Refutability | Judgment |
|---|---|---|---|
| Task fit | P005's specialization was enterprise UI. LP design may be outside specialization | Other participants' primary specializations also were not LP (P001: SaaS, P002: Healthcare, etc.). Task non-fit is not unique to P005 | Largely refutable. However, the distance between enterprise UI and LP may be greater than for others |
| Individual personality traits/motivation | This study did not measure personality traits | Not refutable from observational data. However, P005's verbalization pattern (external attribution style) is consistent with the conceptual alignment failure explanation | Not refutable. Acknowledged as unmeasured confound |
| Conceptual alignment failure (this study's explanation) | P005 recorded multiple verbalizations showing mismatch between AI's and own concept of "LP." Operational modifications consistently failed | Conceptual misalignment confirmed in unit sequence immediately before disengagement, not observed in other successful participants | Most supported explanation in the present data. However, above confounds cannot be excluded |

**P005 disengagement unit sequence (U3→U4→U5):**

**U3:** P005 prompted "Make it a simpler landing page" (L2). AI output: a single-column, spare page. P005: "That's not what I mean by simple. Our company's LPs have much more content" → C-semantic. Here the conceptual mismatch regarding "LP" (P005 assumed a content-rich page; AI assumed a minimal page) became manifest.

**U4:** P005 prompted "Add more content, include service description and pricing table" (L2→L3). AI output: multi-element but scattered layout. P005: "Even after explaining this much, it still doesn't work" → R-surface (superficial dissatisfaction with output). No progression to R-deep.

**U5 (disengagement unit):** P005 prompted "Redo everything. From the beginning" (non-specific instruction). AI output: similar to initial output. P005: "I feel like giving up. I don't even feel like scoring" → disengagement tendency.

*Interpretation:* The U3-U5 sequence demonstrates a pattern in which conceptual misalignment (the meaning gap of "LP") was never resolved while operational details were continuously added, resulting in frustration accumulation and disengagement. Whether this pattern is associated with conceptual alignment failure or with experience differences and AI use experience threshold effects cannot be discriminated from the present data alone.

### *6.5.2 General Limitations*

**Sample size:** N = 5 was designed from the Information Power perspective, but the "disengagement type" (P005) is represented by only one participant. This subcategory has not reached theoretical saturation. Conclusions regarding conceptual alignment are positioned as tentative propositions.

**Tool dependency:** Only Figma Make (version 2025.11) was used. Since the underlying model is not disclosed, replicability with different tools (e.g., Replit AI, Cursor) has not been verified (see Table 6). The impact of output variation was examined in Section 4.3.1, but systematic replicability verification has not been achieved. This study defines replicability not at the level of identical output reproduction but at the level of procedural and analytic replicability—whether SO/ECRT can be re-observed in similar generative AI environments using the same protocol and coding scheme (see Section 4.3). The opacity of the underlying model was managed as a design constraint, and we judged that this constraint does not directly compromise the validity of conclusions on the grounds that the object of analysis is designers' cognitive responses rather than AI output quality. However, the possibility that different model architectures or parameter settings may indirectly influence designers' cognitive processes cannot be excluded and should be verified in future replications.

**Cultural and linguistic context:** All participants prompted in Japanese (see Section 6.4). The association between SO and the high-context nature of Japanese is merely a tendency within the present data scope and is not a causal claim.

**Task dependency:** Limited to the hero section of a B2B SaaS landing page. SO emergence for other design tasks (e.g., dashboard design, mobile app UI) has not been verified.

**Hawthorne effect:** The think-aloud method itself may have increased metacognitive behavior beyond what would occur in natural design practice.

**Limitations of confidence-satisfaction scores:** Scores were used as structured elicitation devices (see Section 3.3) and not for quantitative evaluation. Reliability and validity assessment requires additional verification with larger samples.

**Alternative explanations for SO/ECRT observations:** Alternative explanations exist for this study's central findings (SO, ECRT, the C-semantic and R-deep association) beyond AI's probabilistic nature. First, the influence of task difficulty. The hero section design requires simultaneous integration of multiple dimensions, and this complexity itself may generate trial-and-error patterns resembling SO. However, SO is operationally distinguished from mere trial-and-error by requiring “intentional level descent + verbalization of intentionality” (see Table 2). Second, the influence of time pressure. The 30-minute session constraint may have prompted “efficient exploration” in participants, potentially elevating SO frequency above natural levels. Third, the influence of UI guidance. The UI design of Figma Make itself may have guided specific prompting patterns, raising the possibility that observed SO is a behavioral pattern specific to this tool. Fourth, the influence of individual differences. Differences in years of experience, AI use history, and specialization among the five participants (see Table 4) may confound SO/ECRT emergence patterns. We did not establish control conditions for causal discrimination against the above alternative explanations, and observed patterns are reported as descriptive trends. Causal verification of SO and ECRT requires future confirmatory research including control of tool conditions, manipulation of task difficulty, and stratification by experience level.

**Member checking:** Member checking was not conducted in this study. Analyzed ECRT trajectories and thematic interpretations were not returned to participants for confirmation. This decision was based on two considerations: first, the primary data consisted of real-time think-aloud verbalizations capturing in-the-moment cognitive processes, and retrospective validation risks post-hoc rationalization that could compromise data authenticity (Ericsson & Simon, 1993); second, the abductive coding relied on operational definitions (Table 3) rather than participants’ subjective interpretations. Compensating measures included dual coding by an independent coder ($\kappa = 0.76–0.89$), a priori operational definitions, and proactive negative case analysis (P005). Future replications may incorporate member checking at the thematic level while recognizing its limitations for protocol analysis data.

**Confirmation bias:** The first author is an insider researcher with 25 years of UI/UX design practice experience (see positionality in Section 4.1) and faces the risk of overestimating the utility of the ECRT framework. Three mitigation measures were implemented against this confirmation bias threat. First, dual coding by an independent second coder from outside the HCI domain was implemented, structurally enabling detection of researcher interpretive bias (κ = 0.76–0.82). Second, operational definitions for all codes were finalized a priori (Table 3), minimizing data-driven post hoc modifications. Third, the negative case (P005) inconsistent with the ECRT framework was proactively explored and reported. However, it cannot be asserted that these measures completely eliminated interpretive bias stemming from the insider perspective. Combined with the absence of member checking (see above), independent re-analysis by researchers from outside the design domain is recommended in future replication studies.

**Data auditability:** Raw transcripts are not publicly available due to confidentiality obligations, but anonymized coding matrices (Tables 8 and 9), operational definitions (Table 3), confusion matrices (Table 7), and representative unit excerpts are fully reported within this paper, enabling third-party tracking of the analysis. Additional anonymized data may be requested from the corresponding author (see Section 6.6 for data availability).

### *6.6 Data Availability Statement*

**Shareable materials:** (a) Final codebook—the complete version of operational definitions reported in Table 3 (including judgment criteria and boundary cases). (b) Analysis unit segmentation rules—details of the "1 prompt-1 output pair = 1 unit" segmentation procedure. (c) Anonymized coding inventory—a complete listing of all 34 units with "prompt summary → output summary → key verbalization → assigned codes (both coders)," including representative examples, counterexamples, and disengagement sequences. (d) κ computation script—Python statsmodels-based code-specific κ computation procedures. (e) Unit segmentation judgment log—records of application of consolidation and exclusion criteria (corresponding to Section 4.5 criteria). Materials (a)–(e) contain sufficient information for third parties to trace this study's analytical procedures using the same protocol and coding scheme and to design replications in similar generative AI environments.

**Non-shareable materials:** Raw audio data were deleted in accordance with the deletion policy upon completion of transcription. Screen recording data are stored in encrypted form following anonymization processing.

**Alternative means:** Items (a)–(d) above may be made available as restricted viewing with a non-disclosure agreement (NDA) upon request from the editorial office or reviewers. Additional anonymized data may be requested from the corresponding author within the scope of participant consent.

## 7. Conclusion

### *7.1 Summary*

Through think-aloud protocols with five expert designers ($M$ = 15.4 years of experience), this study found that dialogue with generative AI reorganized externalization not as a unidirectional conversion but as Strategic Oscillation—a dynamic oscillatory movement—within situations where conceptual alignment was established to a certain degree.

### *7.2 Contributions*

**Observational contribution:** Description of the "instruction → consultation" shift and lexical granularity oscillation (Strategic Oscillation). Negative case analysis (P005's disengagement) identified conceptual alignment as a boundary condition.

**Methodological contribution:** Proposal of the ECRT cycle. Comparison with existing reflective models (Table 1) made explicit the theoretical distinctiveness of using probabilistic AI output as the reflection trigger. Systematization of operational definitions (Table 3).

**Design implications:** Presentation of three concrete design patterns (Conceptual Alignment Verification UI, Consultation Shift Nudge, Lexical Granularity Feedback).

### *7.3 Future Directions*

Plans for the second phase include: (1) longitudinal tracking of Strategic Oscillation skill change; (2) expert-novice comparison; (3) additional analysis of disengagement factors; (4) experimental evaluation of design patterns; (5) cross-linguistic comparison (Strategic Oscillation in English prompting environments).

*If tacit knowledge is "knowledge that cannot be verbalized without dialogue with an appropriate other," then AI may serve—precisely because of its unpredictability—as a catalyst for the kind of reflection that human-to-human dialogue alone cannot reach.*

## Disclosure Statement

The author reports there are no competing interests to declare.

## AI Disclosure

In accordance with Taylor & Francis policy, the author discloses the following use of generative AI tools during manuscript preparation. Claude (Anthropic, version Claude Opus 4) was used for language editing and revision support, including grammar correction, phrasing refinement, and formatting of the English-language manuscript. The AI tool was not used for data collection, data analysis, coding, theoretical development, or interpretation of findings. All analytical interpretations, theoretical contributions (ECRT cycle, Strategic Oscillation), empirical findings, and design implications are the sole intellectual work of the author. The author assumes full responsibility for the accuracy and integrity of all content in this manuscript.

## Funding

This research received no specific grant from any funding agency in the public, commercial, or not-for-profit sectors.

## Appendix A. ECRT Subcode Confusion Matrices (All 9 Codes, N = 34 Units)

### A-0. κ Computation Method

For each subcode, a 2 × 2 confusion matrix was constructed based on binary judgments (present/absent) of 34 units. Cohen's κ (Cohen, 1960) was computed as follows:

$P_o = (a + d) / N$ (observed agreement rate: a = both coders judge "present," d = both coders judge "absent")

$P_e = ((a+b)(a+c) + (c+d)(b+d)) / N^2$ (chance agreement rate)

$\kappa = (P_o - P_e) / (1 - P_e)$

Rounding rule: Rounded to two decimal places (third decimal place rounded). Computation was performed using the Python 3.12 statsmodels library and cross-verified by hand calculation.

### A-1. Collision Subcodes

| Coder A  B | C-sem present | C-sem absent | Total (κ = 0.81) |
|---|---|---|---|
| C-sem present | 20 | 1 | 21 |
| C-sem absent | 2 | 11 | 13 |
| **Total** | 22 | 12 | 34 |

| Coder A B | C-syn present | C-syn absent | Total (κ = 0.76) |
|---|---|---|---|
| C-syn present | 15 | 2 | 17 |
| C-syn absent | 2 | 15 | 17 |
| **Total** | 17 | 17 | 34 |

**A-2. Reflection Subcodes**

| Coder A B | R-deep present | R-deep absent | Total (κ = 0.82) |
|---|---|---|---|
| R-deep present | 12 | 1 | 13 |
| R-deep absent | 2 | 19 | 21 |
| **Total** | 14 | 20 | 34 |

| Coder A B | R-surf present | R-surf absent | Total (κ = 0.76) |
|---|---|---|---|
| R-surf present | 14 | 2 | 16 |
| R-surf absent | 2 | 16 | 18 |
| **Total** | 16 | 18 | 34 |

**A-3. Transformation Subcodes**

| Coder A B | T-lex present | T-lex absent | Total (κ = 0.76) |
|---|---|---|---|
| T-lex present | 17 | 2 | 19 |
| T-lex absent | 2 | 13 | 15 |
| **Total** | 19 | 15 | 34 |

| Coder A B | T-str present | T-str absent | Total (κ = 0.81) |
|---|---|---|---|
| T-str present | 11 | 1 | 12 |
| T-str absent | 2 | 20 | 22 |

| Coder A  B | T-str present | T-str absent | Total (κ = 0.81) |
|---|---|---|---|
| **Total** | 13 | 21 | 34 |

| Coder A  B | T-meta present | T-meta absent | Total (κ = 0.76) |
|---|---|---|---|
| T-meta present | 7 | 1 | 8 |
| T-meta absent | 2 | 24 | 26 |
| **Total** | 9 | 25 | 34 |

**A-4. Expectation Subcodes**

| Coder A  B | E-vag present | E-vag absent | Total (κ = 0.76) |
|---|---|---|---|
| E-vag present | 13 | 2 | 15 |
| E-vag absent | 2 | 17 | 19 |
| **Total** | 15 | 19 | 34 |

| Coder A  B | E-spec present | E-spec absent | Total (κ = 0.76) |
|---|---|---|---|
| E-spec present | 17 | 2 | 19 |
| E-spec absent | 2 | 13 | 15 |
| **Total** | 19 | 15 | 34 |

*Note.* Each confusion matrix is based on binary judgments (present/absent) of N = 34 units. Diagonal cells are agreement counts. κ values were computed using Cohen's (1960) formula (see A-0 above). Judgments for each code are independent (multiple codes can be assigned to a single unit); therefore, 2 × 2 confusion matrices were constructed code-by-code. The κ values across all 9 tables are verifiable by recalculation from the above formulas and are fully consistent with the

code-specific κ values reported in the text (Table 7 note). Total agreements across 9 codes = 274, total disagreements = 32 (306 binary judgments).